\documentclass[aoas,MSNbibl,nameyear,dvips]{arximspdf}
\usepackage{multirow}
\usepackage{dcolumn}
\usepackage{graphicx}

\doi{10.1214/11-AOAS491}
\volume{5}
\issue{4}
\pubyear{2011}
\firstpage{2668}
\lastpage{2687}

\makeatletter

\newcolumntype{d}[1]{D{.}{.}{#1}}

\newcommand{\irange}[2]{i = #1,\ldots,#2}
\newcommand{\jrange}[2]{j = #1,\ldots, #2}
\newcommand{\range}[2]{#1,\ldots, #2}

\makeatother

\begin{document}
\begin{frontmatter}

\title{Efficient, adaptive cross-validation for tuning and
comparing models, with application to drug~discovery\thanksref{T1}}
\runtitle{Efficient, adaptive cross-validation}

\thankstext{T1}{Supported by
the Natural Sciences and Engineering Research Council of Canada, the
National Institutes of Health, USA through the NIH Roadmap for Medical
Research, Grant 1 P20 HG003900-01, and the Mathematics of Information
Technology and Complex Systems, Canada.}

\begin{aug}
\author[A]{\fnms{Hui} \snm{Shen}\ead[label=e1]{hshen@interchange.ubc.ca}},
\author[B]{\fnms{William J.} \snm{Welch}\corref{}\ead[label=e2]{will@stat.ubc.ca}}
and~%
\author[C]{\fnms{Jacqueline~M.}~\snm{Hughes-Oliver}\ead[label=e3]{hughesol@stat.ncsu.edu}}

\runauthor{H. Shen, W. J. Welch and J. M. Hughes-Oliver}
\affiliation{University of British Columbia, University of British
Columbia and~North~Carolina State University}
\address[A]{H. Shen\\
School of Population and Public health\\
University of British Columbia\\
Vancouver, BC V6T 1Z3\\
Canada\\
\printead{e1}}
\address[B]{W. J. Welch\\
Department of Statistics\\
University of British Columbia\\
333-6356 Agricultural Road\\
Vancouver, BC V6T 1Z2\\
Canada\\
\printead{e2}}
\address[C]{J. M. Hughes-Oliver\\
Department of Statistics\\
North Carolina State University\\
Raleigh, North Carolina 27695-8203\\
USA\\
\printead{e3}}
\end{aug}

\received{\smonth{12} \syear{2008}}
\revised{\smonth{6} \syear{2011}}

%
\begin{abstract}
\vspace*{12pt}
Cross-validation (CV) is widely used for tuning a model with respect to
user-selected parameters and for selecting a ``best'' model. For
example, the method of $k$-nearest neighbors requires the user to
choose $k$, the number of neighbors, and a neural network has several
tuning parameters controlling the network complexity. Once such
parameters are optimized for a particular data set, the next step is
often to compare the various optimized models and choose the method
with the best predictive performance. Both tuning and model selection
boil down to comparing models, either across different values of the
tuning parameters or across different classes of statistical models
and/or sets of explanatory variables. For multiple large sets of data,
like the PubChem drug discovery cheminformatics data which motivated
this work, reliable CV comparisons are computationally demanding, or
even infeasible. In this paper we develop an efficient sequential
methodology for model comparison based on CV. It also takes into
account the randomness in CV. The number of models is reduced via an
adaptive, multiplicity-adjusted sequential algorithm, where poor
performers are quickly eliminated. By exploiting matching of individual
observations, it is sometimes even possible to establish the
statistically significant inferiority of some models with just one
execution of~CV.
\end{abstract}

%

\end{frontmatter}

\section{Introduction}\label{sect:intro}

The application area that motivated this research illus\-trates the
enormous computational burden that can occur when cross-valida\-tion
(CV) is used to tune and select statistical models. Our
Exploratory~Cen\-ter for Cheminformatics Research,
funded by the National Institutes of
Health Roadmap for Medical Research, is comparing statistical
modeling~me\-thods on assay data from PubChem
(\url{http://pubchem.ncbi.nlm.nih.gov}). For a given assay, activity
(the response variable) against a particular biological target is
measured for thousands or tens of thousands of drug-like
molecules. Several high-dimensional sets of chemical descriptors
(explanatory variables) are available to characterize the chemical
properties of the molecules. A statistical model attempts to
relate biological activity to the chemical descriptors as part of
drug discovery. Currently, for each assay, the web-based
Cheminformatics Modeling Laboratory or ChemModLab
[\citet{HugBroWel2010}]
compares, via CV, 16 statistical methods,
many of which are computationally demanding,
and five candidate sets of descriptor variables. Thus, $16 \times
5 = 80$ modeling strategies are assessed and compared on data sets
with thousands of observations and high-dimensional explanatory
variables.

Moreover, ideally each of these 80 strategies should be tuned with
respect to one or more user-selected parameters, greatly
increasing the number of candidate models to be compared. For
example, a neural network has several user-defined tuning
parameters controlling the network complexity, such as the number
of hidden units and a decay parameter. If many sets of values for
the tuning parameters are tried, potentially hundreds or thousands
of computationally demanding models need to be compared for the
large PubChem data sets.

CV [\citet{Sto1974}] is widely used for this type of study, albeit
usually on a much smaller scale. In a 10-fold cross-validation,
for example, the observations are split into 10 groups or folds,
one group is considered as test data for assessing prediction
accuracy, and the other nine groups are used for model fitting.
This process is repeated with each of the groups in turn as test
data. Thus, further increasing the computational burden already
described, a fixed model (a statistical model with given values of
all tuning parameters and a descriptor set) has to be fitted 10
times.

There is yet another addition to the computational challenge. CV
is based on a random split of the data, and, as we illustrate in
Section \ref{sect:PubChem}, there can be considerable variation
from one split to another. Thus, numerous data splits may be
necessary to compare models reliably.

Thus, the overall computational effort appears to be simply
infeasible for the comprehensive comparisons we have outlined for
large PubChem assay data sets. To our knowledge, currently all
comparisons of this type hence have some degree of unreliability
and/or suboptimality, due to randomness in CV and lack of
effective tuning, respectively.

Much theoretical work has been done on CV.
Stone (\citeyear{Sto1974,Sto1977})
focused mainly on properties for leave-one-out (or $n$-fold) CV.
\citet{Li1987}, \citet{Sha1993} and \citet{Zha1993}
investigated $v$-fold CV
procedures for linear models and general $v$. \citet{Bur1989}
established theoretical results for $v$-fold CV for a~wider class
of models. More recently, \citet{DudVan2005} derived
asymptotic properties for a broad definition of CV (e.g.,
leave-one-out, $v$-fold, Monte Carlo, etc.) for model selection
and performance assessment, and \citet{Yan2006} established the
consistency of CV for classification. The theoretical developments
parallel the extremely wide use of CV by researchers for assessing
and selecting models, for example, \citet{Die1998}, \citet
{HawBasMil2003},
\citet{SinLaa2004b} and \citet{HugBroWel2010}.

In this article we will focus on 10-fold CV, though the
methodology applies to $v$-fold CV for any feasible $v$. We
propose a data-adaptive approach involving multiple repeats of CV
for the candidate models. At any stage, the CV analyses available
from repeated data splits are used to perform a~multiplicity-adjusted statistical test to eliminate all candidate
models that are inferior to at least one other. Only those models
that survive move on to the next stage and have a further CV
performed to increase the test power based on a new, common data
split. In this way, during model tuning, very poor settings of the
tuning parameters are quickly dismissed and computational effort
is concentrated on the best settings. The search terminates when
one setting emerges as the winner, or when the differences in
performance between the surviving settings are practically
unimportant with some statistical confidence. A similar approach
is used to compare optimized models. In the PubChem application
there will be one optimized model for each statistical modeling
strategy, that is, a class of models such as $k$-nearest neighbors
with one of the available descriptor sets. It is also possible to
combine tuning with comparison across optimized models in one
dynamic search.\looseness=-1

Second, we develop more efficient tests for comparing models.
This extends the idea of matching by using the same data splits
across CV analyses [e.g., \citet{Die1998}].
By matching at the level of individual observations rather than data split,
moderate differences in performance between models can sometimes be
detected with just one set of CV analyses from one data split.
Thus, poor performers are potentially eliminated with a minimum of
computing.

Overall, the aim of this article is to develop a sequential
approach for comprehensive and reliable model tuning and selection
via CV. In particular, for the PubChem applications, users of
ChemModLab will have automatic comparison of a vast number of
tuned modeling strategies, with a reasonable turn-around time.

Related to our sequential tests via CV, \citet{MarMoo1997}
developed a ``racing'' algorithm to test a set of models in
parallel. The algorithm sequentially increases data points to
build and test candidate models before using all of the data. In
their paper, leave-one-out CV was used to compute the prediction
error. In contrast, our algorithms use all the data points at all stages,
10-fold CV is implemented to estimate the prediction error, and
computational speed-up is
achieved by reducing the number of models.

The paper is organized as follows. In Section \ref{sect:PubChem}
we describe a typical PubChem data set and the performance
assessment measures relevant to the application. In
Section \ref{sect:variation} we illustrate that there may be
substantial variation in CV performance estimates from one random
data split to another, requiring multiple data splits for reliable
comparison. Section \ref{sect:alg} describes three data-adaptive
algorithms for sequentially comparing models. Whereas
Section \ref{sect:alg} is focused on tuning a given modeling
strategy, that is, a given statistical method and set of data,
Section \ref{sect:compare:methods} considers tuning \textit{and}
comparisons across qualitatively different modeling strategies,
that is, different types of statistical models and/or different
explanatory variable sets. The PubChem data set is used throughout
for illustration. Finally, some conclusions are presented in
Section \ref{sect:conc}.

\section{PubChem AID362 data and assessment measures}\label{sect:PubChem}

ChemModLab\break [\citet{HugBroWel2010}] catalogs the data for five
assays: AID348, AID362, AID364, AID371 and AID377. (``AID''
stands for ``Assay ID.'') In this paper we will focus on AID362,
a formylpeptide receptor ligand binding assay that was conducted
by the New Mexico Molecular Libraries Screening Center; the same
CV comparison methodologies would be applied independently to
other assays in PubChem.

AID362 has assay data for 4,275 molecules. Various responses are
available, but here we work with a binary inactive/active ($0/1$)
measure. Of the 4,275 molecules, only 60 were assayed to be
active. Via computational chemistry, ChemModLab generates five
sets of descriptor (explanatory) variables: Burden numbers,
pharmacophore fingerprints, atom pairs, fragment fingerprints and
Carhart atom pairs, with 24, 121, 395, 597 and 1,578 variables,
respectively.

The purpose of building a statistical model here is to predict the
AID362 inactive/active assay response from the descriptor
variables. Note that the descriptor variables are produced by
\textit{computational} chemistry. Thus, it is feasible to compute them
cheaply for vast numbers of compounds in a chemical library or
even in a virtual library of chemical formulas for molecules that
have not yet been synthesized. The aim of the predictive model,
built from assay data for relatively few molecules, is to choose
the molecules in the bigger library that are most likely to be
active when assayed. Such a~focused search generates ``hits'' for
drug development more efficiently than assaying all the compounds
available, even if this is feasible.

The typical rarity of active compounds and the aim of identifying
a small number of promising compounds in a large\vadjust{\goodbreak} library means
that special predictive performance measures have been developed
for modeling in drug discovery. Misclassification rate, often used
for a binary response, is not appropriate, as even the useless,
null model that always classifies as ``inactive'' will have a high
accuracy rate when active molecules are so rare. The objective is
more to \textit{rank} compounds in terms of their probability of
activity, so that a sample of the desired size of the most
promising compounds can be chosen from a library.

A widely used criterion is a simple function of the number of hits
found, $h_{300}$, among 300 compounds selected using a predictive
model. Specifically, suppose a predictive model generates
$\hat{p}_i$, the probability that compound~$i$ among~$N$ unassayed
compounds is active ($\irange{1}{N}$). We then order the compound
indices via the permutation $\pi$ such that $\hat{p}_{\pi(1)} \ge
\cdots\ge\hat{p}_{\pi(N)}$. Suppose first there are no ties. The
300 compounds indexed by $\pi_{(1)},\ldots, \pi_{(300)}$ are
selected for
assay, and $h_{300}$ is simply the number of actives (hits) found
among them. In general, if $\hat{p}_{\pi(300)}$ ties with the $a +
b$ estimated probabilities
$\hat{p}_{\pi(300-a+1)},\ldots, \hat{p}_{\pi(300+b)}$ for $a \ge
1$ and $b
\ge0$, then $h_{300}$ is defined as
%
%
\begin{equation}\label{eq:h300}
h_{300} = h_{300-a} + \frac{a}{a + b} h_{\mathrm{tie}},
\end{equation}
where $h_{300-a}$ and $h_{\mathrm{tie}}$ are the number of hits found
among the compounds with indices $\range{\pi(1)}{\pi(300-a)}$ and
$\range{\pi(300-a+1)}{\pi(300+b)}$, respectively. This is the
expected number of hits if $a$ compounds are randomly selected
from the $a+b$ with tied probabilities to make a total of 300
selected. No ties for $\hat{p}_{\pi(300)}$ is just a special case
of (\ref{eq:h300}) with $a=1$ and $b=0$.

Initial enhancement (IE), used, for example, by \citet{HugBroWel2010},
is just $(h_{300} / 300) / r$, where $r$ is the
activity rate in the entire collection of~$N$ compounds. Thus, it
measures the rate of finding actives among the 300 chosen
compounds relative to the expected rate under random selection.
A~good model should have IE values much larger than 1. As IE is just
a~linearly increasing function of $h_{300}$, the two criteria are
equivalent, and we use the simpler $h_{300}$ in this article.
Users concerned about the arbitrariness of selecting 300 compounds
may prefer the average hit rate (AHR) proposed by \citet{Wan2005},
which averages performance over all selection sizes but favors
models which rank active compounds ahead of inactive compounds in
terms of $\hat{p}_i$. Algorithms 1 and 3 in Section \ref{sect:alg}
could be applied directly to AHR without modification.

In defining the assessment measure $h_{300}$, we have assumed
there is a~training data set to build a model and a further
independent test set of $N$ compounds available to assess it. This
article is concerned with CV, however, where the same $n$
observations are used for training and for testing. Under 10-fold
CV, for instance, when a particular data fold is removed to serve
as test data, the model fitted to the remaining data generates the
$\hat{p}_i$ values for the compounds in that fold. After cycling
through all 10 folds, the $\hat{p}_i$ values are put\vadjust{\goodbreak} together so
that there is a $\hat{p}_i$ for all $n$ compounds. We then define
$h_{300}$ (or an alternative criterion) exactly as above except
that we choose 300 compounds from the $n \ge300$ instead of from
an independent set of size $N$.

\section{Variation in cross-validation}\label{sect:variation}

We now demonstrate that there can be substantial variation in the
performance estimates from 10-fold CV from one random split of the
data to another, potentially requiring multiple splits for
reliable model tuning or selection. For illustration, the PubChem
AID362 assay data will be modeled using a neural network (NN)
[see, e.g., \citet{Rip1996}] with one hidden layer and a
variation of
Burden numbers [\citet{Burden1989}] as the descriptor set. For the
AID362 assay, there are 60 active compounds among 4,275 molecules,
and the Burden number descriptor set has 24 variables.

We will tune two important parameters of the NN: the number of
units in the hidden layer, which controls the size or complexity
of the network, and a decay parameter, where smaller values shrink
the network weights less and lead again to a more complex network.
In this tuning study, size takes the values 5, 7 and~9, and decay
takes the values 0.1, 0.01 and 0.001. Thus, tuning will select
among $3 \times3 = 9$ models generated by all combinations of
the two tuning parameters.

%
%
\begin{figure}

\includegraphics{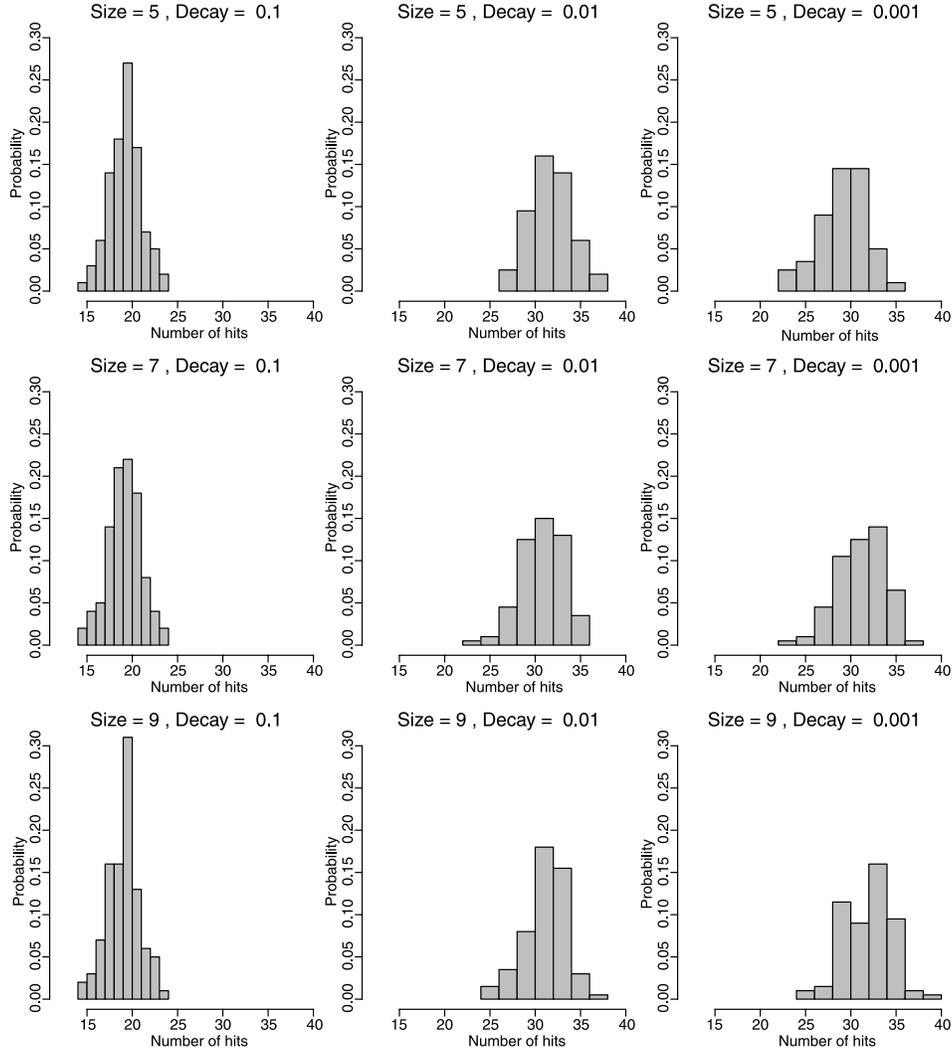}

\caption{Histograms showing the distribution of $h_{300}$ values
from 10-fold CV across 100 data splits for neural networks with
different values of size and decay, and Burden numbers as the
descriptor set.} \label{fig:hist:nn:burden}
\end{figure}

For each model, 10-fold CV is run for 100 random splits of the
data, and the histograms in Figure \ref{fig:hist:nn:burden} show
the estimated distributions of the $h_{300}$ assessment measure
defined in (\ref{eq:h300}). We can see that there are considerable
differences between the $h_{300}$ distributions across the tuning
parameters values considered, that is, tuning is important. There is
also considerable variation within a fixed set of tuning parameter
values. For example, for $\mathrm{size}=5$ and $\mathrm{decay}=0.01$,
which is one of
the better performing models, $h_{300}$ ranges from 26 to 37. We
will take the population mean performance over a large number of
repeated cross-validations as a reliable measure of performance,
reliable in the sense that random cross-validation variation is
eliminated. Table \ref{tab:nn:burden} displays the observed sample
means of $h_{300}$ with their standard errors. Models~2, 5, 6, 8
and 9 have better sample means than models 1, 3, 4 and 7.
Moreover, the standard errors are fairly small relative to the
differences between the sample means across these two groups,
suggesting that the weaker performers could be dismissed with
fewer than 100 random data splits, whereas finding the best
parameter values among the better models will take considerable
work (though perhaps not requiring 100 random data splits). This
is the basic idea underlying the adaptive algorithms of
Section~\ref{sect:alg}.\looseness=-1

%
%
\begin{table}
\caption{Sample means of $h_{300}$ for 10-fold CV across 100 data
splits for neural networks with different values of size and
decay, and Burden numbers as the descriptor set, applied to the
PubChem \textit{AID362} assay data} \label{tab:nn:burden}
\begin{tabular*}{\tablewidth}{@{\extracolsep{\fill}}ld{2.2}d{1.2}d{2.3}d{2.2}
d{2.2}d{2.3}d{2.2}d{2.2}d{2.3}@{}}
\hline
& \multicolumn{9}{c@{}}{\textbf{Neural network model}} \\[-4pt]
& \multicolumn{9}{c@{}}{\hrulefill}\\
& \multicolumn{1}{c}{\textbf{1}} & \multicolumn{1}{c}{\textbf{2}}
& \multicolumn{1}{c}{\textbf{3}} & \multicolumn{1}{c}{\textbf{4}}
& \multicolumn{1}{c}{\textbf{5}} & \multicolumn{1}{c}{\textbf{6}}
& \multicolumn{1}{c}{\textbf{7}} & \multicolumn{1}{c}{\textbf{8}}
& \multicolumn{1}{c@{}}{\textbf{9}} \\
\hline
Size & 5 & 5 & 5 & 7 & 7 & 7 & 9 & 9 & 9 \\
Decay & 0.1 & 0.01 & 0.001 & 0.1 & 0.01 & 0.001 & 0.1 & 0.01 & 0.001 \\
\# of hits & 18.7 & 31.2 & 28.7 & 18.6 & 30.2 & 30.7 & 18.5 & 30.7 &
31.4 \\
S.E. & 0.18 & 0.24 & 0.27 & 0.19 & 0.24 & 0.27 & 0.18 & 0.22 & 0.26 \\
Rank & 7 & 2 & 6 & 8 & 5 & 3 & 9 & 4 & 1 \\
\hline
\end{tabular*}
\end{table}

\begin{figure}[b]

\includegraphics{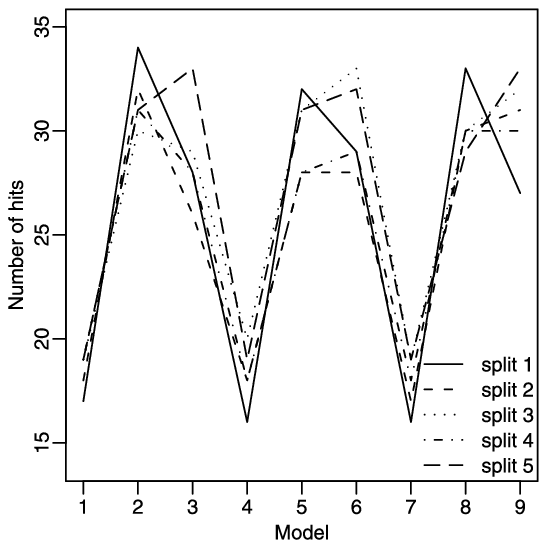}

\caption{$h_{300}$ values for 10-fold CV and five data splits
(five lines) for neural networks with different values of size and
decay, and Burden numbers as the descriptor set.}
\label{fig:blocks:nn:burden}
\end{figure}

Such a comparison should take into account that data split would
naturally be a blocking factor. Every time a random data split is
generated, all models under consideration are assessed via CV
using this same split. Thus, the 100 data splits leading to the
data in Figure \ref{fig:hist:nn:burden} are 100 blocks.
Figure~\ref{fig:blocks:nn:burden} shows the results for five
%
%
%
blocks, with each line representing one split. The approximate
parallelism of the curves indicates that including split as a
blocking factor will lead to more powerful comparisons.\vadjust{\goodbreak}
Figure \ref{fig:blocks:nn:burden} also suggests that comparing
models based on just one split may lead to a biased estimator of
performance. For each curve, suppose we select the model (set of
tuning parameter values) with the largest observed value of
$h_{300}$. We note first that, probably due to selection bias, the
$h_{300}$ value of the winning model tends to be in the upper tail
of its distribution in Figure \ref{fig:hist:nn:burden}. Second,
for the fifth split, suboptimal model 3 has the best value of
$h_{300}$. Thus, there is a~need for multiple splits for reliable
assessment and comparison.\vadjust{\goodbreak}

\section{Algorithms for adaptive model search via sequential CV}\label
{sect:alg}

\subsection{Algorithm 1 (data splits as blocks)}\label{sect:alg:split}

Suppose there are $m$ models to be compared. For much of this
article, we will be comparing $m$ sets of values for the tuning
parameters of a given type of statistical modeling method, in the
context of a fixed descriptor (explanatory variable) set.
Comparisons across qualitatively different statistical models
and/or different sets of explanatory variables are also possible,
however (Section \ref{sect:compare:methods}). The algorithm will
attempt to remove models sequentially until $m$ is reduced to 1.

At each iteration, a new random data split is created for CV,
and 10-fold (or $v$-fold in general) CV estimates of
performance are computed for the surviving models.
For each model, CV requires 10 model fits for the new split.
Thus, regular CV is applied;
the various algorithms to be described are efficient by
reducing the number of times such a regular CV analysis has to be performed.

Specifically,
suppose there are $m$ surviving models, and results from running
10-fold CV are available for $s \geq2$ random splits. The
assessment measure is computed for every model and split. We will
use $h_{300}$ in (\ref{eq:h300}), but for this first version of
the algorithm any user-defined measure could be employed, for example,
the average hit rate in Section \ref{sect:PubChem} or, for a
continuous response, the empirical predictive mean squared error. In
general, $y_{ij}$ will denote the CV assessment measure for model $i$
and data split $j$.

If a randomly chosen split is applied across all models, split is
a blocking factor, and we can model $y_{ij}$ as generated by
\[
Y_{ij} = \mu+ \tau_i + \beta_j + \varepsilon_{ij} \qquad
(i=1,\ldots
,m; j=1,\ldots,s),
\]
where $\mu$ is an overall effect, $\tau_i$ is the effect of model
$i$, $\beta_j$ is the effect of split~$j$, and $\varepsilon_{ij}$
for $i=1,\ldots,m$ and $j=1,\ldots,s$ are random errors, assumed
to have independent normal distributions with mean $0$ and
variance $\sigma^2$. This is the model for a randomized block
design, though we point out that randomization within a block, for
example, executing the analyses for the $m$ models in a random
order, has no relevance for such a ``computer experiment.''

We want to test the hypotheses
\[
H_0\dvtx\tau_i=\tau_{i'}\quad\mbox{versus}\quad H_1\dvtx\tau
_i\neq\tau_{i'}
\]
for all $i\neq i'$. For a particular pair of models indexed by $i$
and $i'$, rejecting $H_0$ in favor of $H_1$ at some significance
level implies that one of the models may be eliminated as inferior
to the other. After removing all such dominated models, at the
next iteration, further CV computational effort will be
concentrated on the surviving models.

At least initially, $m$ may be large, and a multiplicity-adjusted
test is desirable.
Tukey's test [\citet{DeaVos1999}, Chapter 4, \citet{Mon1997},
Chapter 3]
is a common choice for such multiple comparisons,
and we adopt it throughout.
Other tests for multiple comparisons could be
applied, such as Fisher's least significant difference test or
Duncan's multiple range test, etc. [\citet{Mon1997}, Chapter 3].
Let
%
%
\begin{equation}\label{eq:ybar}
\bar y_{i\cdot}= \frac{1}{s}\sum_{j=1}^s y_{ij}
\end{equation}
denote the sample mean performance over the $s$ splits for model
$i$ ($i=1,\ldots, m$). For any $i\ne i'$, the null hypothesis
$H_0$ is rejected in favor of $H_1$ at level $\alpha$ if
\[
| \bar y_{i\cdot}- \bar y_{i'\cdot}| > T_{\alpha}(m,s),
\]
where
%
%
\begin{equation}\label{eq:tukey:value}
T_{\alpha}(m,s)=q_{\alpha}\bigl(m,(m-1)(s-1)\bigr)\sqrt{ \frac{
\operatorname{MSE}(m,s)}{s}}\vadjust{\goodbreak}
\end{equation}
is the Tukey value,
$q_{\alpha}(m,(m-1)(s-1))$ is the studentized range
statistic with $m$ and $(m-1)(s-1)$ degrees of freedom, and
$\operatorname{MSE}(m,s)$ is the mean square for error under a
randomized-block analysis of variance with $m$ models (treatments)
and $s$ splits (blocks).
A set of simultaneous $100(1-\alpha)$
percent confidence intervals for all pairwise differences
$\tau_i-\tau_{i'}$ for $i\neq i'$ is given by
%
%
\begin{equation}\label{eq:CI}
\tau_i-\tau_{i'}\in\bigl(\bar y_{i\cdot}-\bar y_{i'\cdot} \pm
T_{\alpha}(m,s)\bigr).
\end{equation}

The properties of statistical tests in analysis of variance models
in general are often justified via randomization
[e.g., Kempthorne (\citeyear{Kem1952,Kem1955})].
As already noted, randomization of models to a split (block) is irrelevant
here,
and it is questionable whether the nominal significance level $\alpha$
is actually achieved under the null hypothesis.
In any case, as the algorithm iterates and more blocks are added,
a sequence of tests is performed.
Even if each stage has the correct significance level for removing
a model when it is not inferior,
the entire procedure would not.
Overall, then, $\alpha$ is best viewed as controlling a greedy algorithm,
where larger values would remove models more aggressively,
and the gain in computational speed is accompanied by
more risk of converging to a suboptimal model.
We use $\alpha= 0.05$ throughout for empirical demonstrations
and compare the solutions found with more exhaustive searches.

%
%
\begin{figure}
%
%
\fbox{\begin{tabular}{c}

\includegraphics{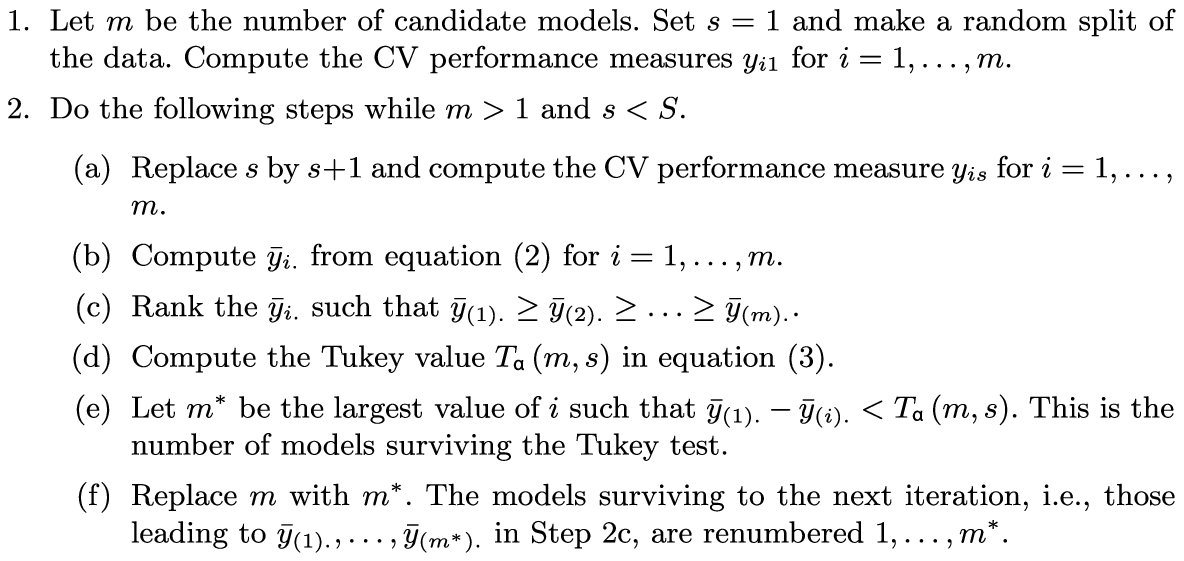}

\end{tabular}}
\caption{Adaptive model search via sequential CV (Algorithm 1:
iterate until one model is left or a maximum of $S$ data splits has
been performed).} \label{fig:alg1:code}
\end{figure}

Figure \ref{fig:alg1:code} gives pseudo code for the above
sequential algorithm.
It iterates until only one model is left, subject to a maximum of
$S$ random data splits and hence $S$ CV analyses for any model. We
use $S=100$ hereafter. Note that this algorithm needs at least two
executions of CV for each initial model from two random data
splits.

For illustration, we revisit the PubChem AID362 example in
Section \ref{sect:variation}, where the descriptor set is formed
from Burden numbers, and the problem is to tune the parameters
decay and size for a neural network model. The nine candidate
models, that is, the nine combinations of decay and size values, were
given in Table \ref{tab:nn:burden}.

Table \ref{tab:alg:split} shows the results of applying
Algorithm 1 to this example.
%
%
\begin{table}
\caption{Algorithm 1 applied to tuning the values of size and
decay for a neural network for the PubChem \textit{AID362} assay data with
Burden numbers as the descriptor set. The models surviving after
each split are denoted by a check mark} \label{tab:alg:split}
\begin{tabular*}{\tablewidth}{@{\extracolsep{\fill}}lccccccccc@{}}
\hline
\multirow{2}{37pt}[-7pt]{\textbf{Number of splits}}
& \multicolumn{9}{c@{}}{\textbf{Neural network model}}
\\[-4pt]
& \multicolumn{9}{c@{}}{\hrulefill}\\
& \textbf{1} & \textbf{2} & \textbf{3} & \textbf{4} & \textbf{5}
& \textbf{6} & \textbf{7} & \textbf{8} & \textbf{9} \\
\hline
\hphantom{00}0 & \checkmark&\checkmark&\checkmark& \checkmark
&\checkmark
&\checkmark&\checkmark&\checkmark&\checkmark\\
\hphantom{00}2 & &\checkmark&\checkmark& &\checkmark&\checkmark&
&\checkmark
&\checkmark\\
\hphantom{00}3 & &\checkmark&\checkmark& &\checkmark&\checkmark&
&\checkmark
&\checkmark\\
\hphantom{00}4 & &\checkmark&\checkmark& &\checkmark&\checkmark&
&\checkmark
&\checkmark\\
\hphantom{00}5 & &\checkmark& & &\checkmark&\checkmark& &\checkmark
&\checkmark
\\
\hphantom{00}$\vdots$& $\vdots$ & $\vdots$ & $\vdots$ & $\vdots$
& $\vdots$ &
$\vdots$ & $\vdots$ & $\vdots$ & $\vdots$ \\
\hphantom{0}57 & &\checkmark& & &\checkmark&\checkmark& &\checkmark
&\checkmark\\
\hphantom{0}58 & &\checkmark& & & &\checkmark& &\checkmark
&\checkmark\\
\hphantom{00}$\vdots$& $\vdots$ & $\vdots$ & $\vdots$ & $\vdots$
& $\vdots$ & $\vdots
$ & $\vdots$ & $\vdots$ & $\vdots$ \\
\hphantom{0}68 & &\checkmark& & & &\checkmark& &\checkmark
&\checkmark\\
\hphantom{0}69 & &\checkmark& & & & & &\checkmark&\checkmark\\
\hphantom{0}70 & &\checkmark& & & & & & &\checkmark\\
$\vdots$& $\vdots$ & $\vdots$ & $\vdots$ & $\vdots$ & $\vdots$ &
$\vdots$ & $\vdots$ & $\vdots$ & $\vdots$ \\
100 & &\checkmark& & & & & & &\checkmark\\
\hline
\end{tabular*}
\end{table}
After $s=2$ splits, the average $h_{300}$ values for the nine
models, $\bar{y}_{1},\ldots, \bar{y}_{9}$, are
\[
17.5, 33.0, 27.0, 17.0, 30.0, 28.5, 16.5, 31.5, 29.0,
\]
$\operatorname{MSE}(9,2) = 3.39$, and the Tukey value is 7.51 for
significance level $\alpha= 0.05$. Since $\bar y_{2.}-\bar
y_{i\cdot}>7.51$ for $i = 1$, 4 and 7, these three models are
dismissed. Recall from Table \ref{tab:nn:burden} that they are indeed
the worst when averaged over 100 splits, but the sequential algorithm
eliminates them after just two CV splits. Hence, in
Table~\ref{tab:alg:split}, only models 2, 3, 5, 6, 8 and 9 survive the
second split and are included for a third round of CV based on another
split. After five splits, model 3 is removed. Models 5, 6 and 8 are
removed after 58, 69 and 70 splits, respectively. The two remaining
models, 2 and 9, are still in contention when the algorithm stops due
to restricting the computational effort to 100 splits. From the average
$h_{300}$ values given in Table~\ref{tab:nn:burden}, we
know that these two models are very similar in
performance, and are hard to distinguish. This motivates Algorithm 3 in
Section \ref{sect:alg:stopping}, but next we improve
Algorithm~1.\vadjust{\goodbreak}

\subsection{Algorithm 2 (observations as blocks)}\label{sect:alg:active}

Algorithm 1 in Section \ref{sect:alg:split} needs at least two CV
data splits for every one of the $m$ models, which may be
computationally expensive if $m$ is large. We now describe another
multiplicity-adjusted test, aimed at eliminating bad models after
only one CV data split.

Unlike Algorithm 1, the revised algorithm is applicable only to an
assessment measure that is a sum or average of contributions from
individual observations. The criterion $h_{300}$
in (\ref{eq:h300}) is of this form, and we continue to use it, but
we note that only the active compounds in the data set can make a
nonzero contribution to $h_{300}$, and it is sufficient to
consider them only. Specifically, suppose there are $A \ge2$
active compounds in the data set ($A = 60$ for the AID362 assay).
For any given model, its CV analysis leads to estimated
probabilities of activity $\hat{p}_{\pi(1)} \ge\cdots\ge
\hat{p}_{\pi(n)}$ for the $n$ compounds in the data set. We can
write
\[
h_{300} = \sum_{j=1}^{A} y_j^*,
\]
where $y_j^*$ is the contribution from active compound $j$. From
the definition of $h_{300}$ in (\ref{eq:h300}),
%
%
\begin{equation}\label{eq:y:contrib}
y_j^* = \cases{
1, &\quad if active compound $j$ is one of the first\cr
&\quad$300-a$ compounds selected, \vspace*{1pt}\cr
\dfrac{a}{a+b}, &\quad if active compound $j$ appears
among the\cr
&\quad$a+b$ compounds with $\hat{p}$ tying with $\hat{p}_{\pi
(300)}$,\vspace*{1pt}\cr
0, &\quad otherwise.}
\end{equation}
(Recall that $\hat{p}_{\pi(300)}$ ties with the $a + b$ estimated
probabilities $\hat{p}_{\pi(300-a+1)},\ldots,\allowbreak \hat{p}_{\pi
(300+b)}$ with $a
\ge1$ and $b \ge0$, which includes no ties if $a=1$ and $b=0$.)

For example, suppose a CV analysis leads to estimated
probabilities of activity $\hat{p}_{\pi(1)} \ge\cdots\ge
\hat{p}_{\pi(n)}$ such that $\hat{p}_{\pi(300)}$ has the eight
ties $\hat{p}_{\pi(298)},\ldots, \hat{p}_{\pi(305)}$. Of the, say, $A=60$
active compounds, 25 have estimated probabilities among
$\hat{p}_{\pi(1)},\ldots, \hat{p}_{\pi(297)}$; they each have
$y_j^* = 1$
in (\ref{eq:y:contrib}) because they must each contribute one hit
to $h_{300}$. Another two active compounds have estimated
probabilities among $\hat{p}_{\pi(298)},\ldots, \hat{p}_{\pi
(305)}$; they
each have $y_j^* = 3 / 8$, the probability of being selected
298th, 299th or 300th when the eight selections
$\range{298}{305}$ are made in random order.

We now consider CV to compare $m$ models based on one
common random data split. Let $y_{ij}^*$ be the contribution of
active compound $j$ to $h_{300}$ for model~$i$,
for $\irange{1}{m}$ and $\jrange{1}{A}$. A
multiplicity-adjusted test parallels that in
Section \ref{sect:alg:split}. In the randomized-block analysis,
the blocks are now the $A$ active compounds rather than data
splits (there is only one). If a randomly chosen split is applied
across all models, we can model $y_{ij}^*$ as generated by
\[
Y_{ij}^* = \mu^* + \tau^*_i + \beta^*_j + \varepsilon_{ij}^*
\qquad(i=1,\ldots,m; j=1,\ldots,A),\vadjust{\goodbreak}
\]
where $\mu^*$ is an overall effect, $\tau^*_i$ is the effect of
model $i$, $\beta^*_j$ is the effect of active compound $j$,
and the $\varepsilon_{ij}^*$ are
random errors, assumed to have independent normal distributions
with mean $0$ and variance $\sigma^{*^2}$. Similarly, the Tukey
value in (\ref{eq:tukey:value}) is replaced by
\[
T_{\alpha}(m, A) = q_{\alpha}\bigl(m,(m-1)(A-1)\bigr)\sqrt{ \frac
{\operatorname{MSE}(m,A)}{A}},
\]
where the studentized range statistic $q$ has degrees of freedom
$m$ and $(m-1)(A-1)$. Analogous hypothesis tests eliminate all
models significantly different from the one with the best observed
performance.

If $s \ge2$ data splits have been made, we could define blocks in
terms of active compounds \textit{and} data splits, that is, the
number of blocks would be $sA$. Some experimentation indicates
that Algorithm 1 in Section \ref{sect:alg:split} eliminates
inferior models faster, however, for $s \ge2$. Thus, for
Algorithm 2 we use the Tukey test based on active compounds as
blocks only for the first data split. After very poor models are
eliminated, a second split is made and the data-adaptive search
proceeds for the surviving models as in
Section \ref{sect:alg:split} with $s \ge2$ splits as blocks.

\begin{figure}

\includegraphics{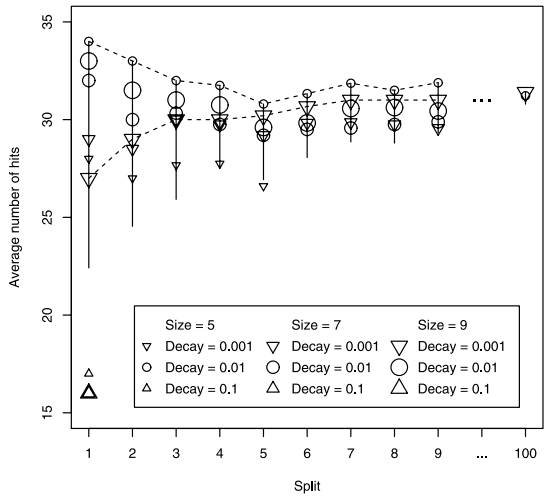}
\caption{Algorithm 2 applied to tuning a neural network (NN) for the
PubChem AID362 assay data with Burden numbers as the descriptor set.
NN sizes of 5, 7 and 9 are denoted by
small, medium and large plotting symbols, respectively,
and decay values of 0.001, 0.01 and 0.1 are denoted by
$\triangledown$, $\circ$~and~$\vartriangle$, respectively.
The two models surviving after 100 splits
are connected with dashed lines.} \label{fig:alg2}
\end{figure}

We now revisit the example of tuning a neural network in
Section \ref{sect:alg:split} (the results for Algorithm 1 were
presented in Table \ref{tab:alg:split}). Figure \ref{fig:alg2}
depicts Algorithm 2's progress, plotting $h_{300}$ versus split.
%
%
%
After running one split of 10-fold CV, the model with $\mbox{size} = 5$ and
$\mbox{decay} = 0.01$ has the largest $h_{300}$ value. The vertical line
drawn down from this value has length $T_\alpha(m,A)$, where $m=9$
and $A=60$. It is based on Tukey's test with the 60 active
compounds as blocks. The $h_{300}$ values for models 1, 4 and 7
fall below this line and they are eliminated with one CV split.
For $2, 3, \ldots$ splits, $\bar y_{i\cdot}$, the average of $h_{300}$
over the splits, is plotted for the surviving models, and the
vertical lines have length $T_\alpha(m,s)$, where $m$ is the
number of models surviving to $s$ splits. It is seen that after 2,
3 or 4 splits of CV, no further models are eliminated. After 5
splits, model 3 is dismissed, and after 9 splits models 2, 5, 6, 8
and 9 still survive. The two models with the largest~$h_{300}$
averages, models 2 and 9, are connected with dashed lines in
Figure~\ref{fig:alg2}. These models are still competitors after
100 splits of CV. The vertical line drawn at 100 splits is very
short; nonetheless, these models are so close in performance that
they cannot be distinguished. Again, this motivates
Algorithm~3.\looseness=1

\subsection{Algorithm 3 (modified stopping criterion)}\label
{sect:alg:stopping}

As has already been illustrated, if the predictive performances
for several candidate models are very similar, it can take many
data splits and CV analyses to distinguish them. Particularly for
model tuning, it would be more efficient to modify the stopping
criterion so that the algorithm stops once it is clear that the
current leading performer cannot be beaten by a practically
important amount.\looseness=1

We implement such a stopping criterion via the confidence
intervals in~(\ref{eq:CI}). Again, rank the $m > 1$ models surviving
at any iteration in terms of their average predictive
performances, that is, $\bar y_{(1)\cdot}\geq\bar y_{(2)\cdot}\geq
\cdots\geq\bar y_{(m)\cdot}$. Notationally, we will use $s$ (number
of splits) for the number of blocks in these averages, but the
same method can be applied with observations as blocks as in
Section \ref{sect:alg:active}. From~(\ref{eq:CI}),
\[
\tau_{(i)} - \tau_{(1)} \in\bigl(\bar y_{(i)\cdot}-\bar y_{(1)\cdot}
\pm T_{\alpha}(m,s)\bigr)\qquad
\mbox{for $\irange{2}{m}$}.
\]
At some confidence level, we want
to be sure that $\tau_{(i)} - \tau_{(1)} < p_0$ for all
$\irange{2}{m}$, where $p_0$ is a given practically insignificant
performance difference.
Thus, to stop with the model giving $\bar y_{(1)\cdot}$ declared as
the winner,
$\bar y_{(i)\cdot}-\bar y_{(1)\cdot}\allowbreak + T_{\alpha}(m,s) < p_0$ for
all $\irange{2}{m}$.
As the $\bar y_{(i)\cdot}$ are nonincreasing with~$i$,
the revised stopping criterion is simply
%
%
\begin{equation}\label{eq:p0}
\bar y_{(2)\cdot} - \bar y_{(1)\cdot} + T_{\alpha}(m,s) < p_0.
\end{equation}
%

\begin{figure}

\includegraphics{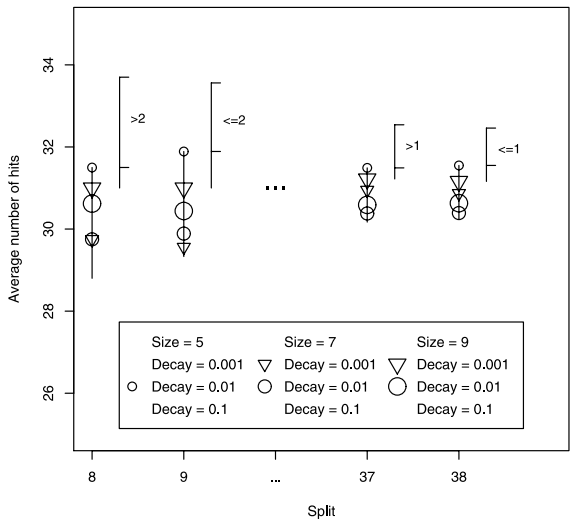}

\caption{Algorithm 3 applied to tuning a neural network (NN) for the
PubChem AID362 assay data with Burden numbers as the descriptor set.
NN sizes of 5, 7 and 9 are denoted by
small, medium and large plotting symbols, respectively,
and decay values of 0.001 and 0.01 are denoted by
$\triangledown$ and $\circ$, respectively.
The results for splits 1--7 are as in Figure \protect\ref{fig:alg2}
and are not shown.
All NNs with decay of 0.1 have been eliminated by split 8,
as is the NN with size of 5 and decay of 0.001.}
\label{fig:alg3}
\end{figure}

For the example of tuning a neural network for the AID362 assay
data and Burden number descriptors,
the values of $\bar y_{(2)\cdot} - \bar y_{(1)\cdot} + T_{\alpha
}(m,s)$ in (\ref
{eq:p0}) for data splits
1--38 are as follows:
\[
10.59, 6.96, 5.08, 3.16, 3.27, 2.60, 2.14, 2.20, 1.67, \ldots, 1.05, 0.91.
\]
(The hybrid observations/splits as blocks algorithm of
Section \ref{sect:alg:active} is being used here.) If\vadjust{\goodbreak} we take $p_0
= 1$ as the practically insignificant performance difference, the
algorithm stops after 38 splits of 10-fold CV, with surviving
models 2, 5, 6, 8 and~9. Model 2 with $\mbox{size} = 5$ and $\mbox
{decay} = 0.01$
would be declared the ``tuned'' model for practical purposes. If
we set $p_0 = 2$, the algorithm stops after just nine splits.
Models 2, 5, 6, 8 and 9 are again the survivors, and again model 2
is declared the winner for the neural networks/Burden numbers
modeling strategy. Figure \ref{fig:alg3} illustrates the
iterations of the algorithm. In particular, the vertical lines
shown to the right of the performance averages for 8, 9, 37 and
38 splits start at $\bar y_{(2)\cdot}$ and have length
$T_{\alpha}(m,s)$.
If they extend less than $p_0$ past $\bar{y}_{(1)\cdot}$
[i.e., $\bar y_{(2)\cdot} + T_{\alpha}(m,s) < \bar y_{(1)\cdot} + p_0$],
then the revised stopping criterion (\ref{eq:p0}) is
satisfied.

%
%

Recall that when we try to establish the one winning model via
Algorithms 1 or 2, 100 data splits and CV analyses are
insufficient to separate models 2 and~9. Therefore, the modified
stopping criterion saves considerable computing time here.

\section{Comparing statistical methods or explanatory variable
sets}\label{sect:compare:methods}

Recall that Section \ref{sect:intro} described 80 statistical
methods/descriptor set modeling strategies compared by
ChemModlab. When comparing qualitatively different statistical
methods and/or explanatory variable sets, there are two possible
search implementations:
\begin{itemize}
\item\textit{Tune then compare}:
\begin{itemize}[Step 1.]
\item[Step 1.] Tune each modeling strategy independently by
repeating one of the algorithms in Section \ref{sect:alg}. For
ChemModLab, this would mean 80 tuning searches.\vadjust{\goodbreak}
\item[Step 2.]
Compare the tuned models, again by applying one of the algorithms
in Section \ref{sect:alg}.
\end{itemize}

As we shall illustrate, the CV analyses in Step 1 can be
reused in Step~2, possibly leading to minimal further computing
at Step 2. This approach is preferred when one wants to assess the
performance of every modeling strategy after tuning. It requires
many searches in Step 1, however.
\item\textit{Simultaneously tune
and compare}: Carry out one search, simultaneously tuning and
comparing the model strategies.

This approach, we shall see, can require much less computing. Its
drawback, however, is that it does not necessarily provide
accurate estimation of the predictive performances of suboptimal
strategies; we just infer they are dominated by the winning
strategy.
\end{itemize}

For simplicity, we will illustrate these two search
implementations by comparing two statistical methods/descriptor
sets for the AID362 PubChem assay. Extension to all 80 strategies
explored by ChemModLab is straightforward. One strategy is a
neural network with Burden number descriptors, which we call
NN/Burden. NN/Burden was investigated in Section \ref{sect:alg},
and we already know that $\mbox{size}=5$ and $\mbox{decay}=0.01$ provides good
values of the tuning parameters. The second strategy is
$k$-nearest neighbors with Carhart atom pairs as explanatory
variables, which we call KNN/Carhart.

\begin{figure}

\includegraphics{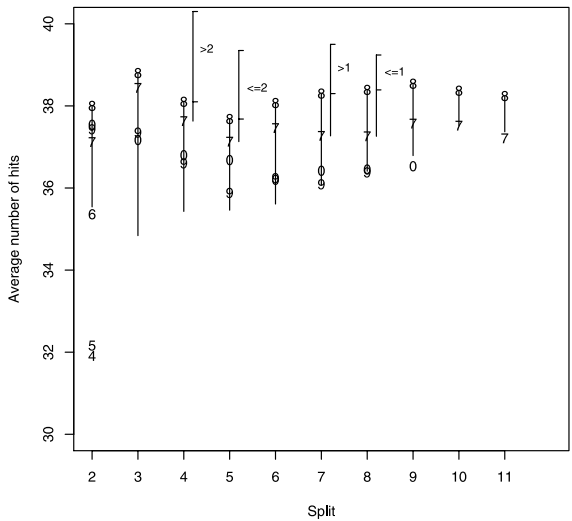}

\caption{Algorithms 2 and 3 applied to tuning $k$-nearest neighbors
(KNN) for the PubChem \textit{AID362} assay data with Carhart atom
pairs as the descriptor set. Values of $k=4,\ldots,10$ are denoted by
the plotting symbols $4,\ldots,9,0$.}
\label{fig:KNN:Carhart}
\end{figure}

For KNN/Carhart, we need to tune $k$, the number of neighbors.
We consider $k$ in the range $1, 2, \ldots, 10$.
Figure \ref{fig:KNN:Carhart} shows the results of running
Algorithms 2 and 3 in Sections \ref{sect:alg:active}
and \ref{sect:alg:stopping}. Algorithm 2 stops after 11 CV data splits,
and the model with $k=8$ emerges as the winner.
(This agrees with more exhaustive computations to check our
algorithm.) If we use the stopping criterion $p_0 = 1$
in (\ref{eq:p0}), the algorithm stops after just eight data
splits. With $p_0 = 2$, only five data splits are required. All
these variants point to $k=8$.

%
%

We now consider the tune-then-compare implementation for comparing
NN/ Burden with KNN/Carhart. For definiteness, we take $p_0 = 2$
in (\ref{eq:p0}) as the stopping criterion. In Step 1, the two
strategies are tuned independently, which has already been
described. In Step 2, NN/Burden ($\mbox{size}=5$ and $\mbox{decay}=0.01$) is
compared with KNN/Carhart ($k=8$). Running Algorithm 3 in
Section \ref{sect:alg:stopping} for three data splits is
sufficient to establish that tuned KNN/Carhart is better than
tuned NN/Burden at significance level 0.05. The $95\%$ confidence
interval for the difference in mean $h_{300}$ is $[2.31, 11.29]$.
In Step 1, the total number of model fits
(with 10 fits per 10-fold CV) is
$10 (9 + 6\times4 + 5\times4) = 530$ for NN/Burden and
$10(10+7+4\times3)=290$ for
KNN/Carhart. The same data splits were used for NN/Burden and
KNN/Carhart. Thus, for Step 2, the first splits from
Step 1 can be reused and no further CV computations are required.
Therefore, the total number of model fits in
both steps to establish that KNN/Carhart (with $k=8$) is superior
is $530+290=820$.
No model required more than 10 random splits to define the CV folds.

For the simultaneous tune and compare implementation, the nine
NN/\break Burden models (with different values of size and decay) and the
10 KNN/\break Carhart models (with different values of $k$) are put
together as $m=19$ initial models. The results of running
Algorithm 3 in Section \ref{sect:alg:stopping} with $p_0 = 2$ are
shown in Table \ref{tab:one:search}. We see after just one split,
with the active compounds as blocks, three NN/Burden models and
one KNN/Carhart model are eliminated. After two data splits, with
splits as blocks, only one NN/Burden model and five KNN/Carhart
models survive. After three splits, the remaining NN/Burden model
is eliminated. The five KNN/Carhart models left survive through
five splits, when the stopping criterion is satisfied. These
models have $k = 6, 7, 8, 9$ and 10 and average $h_{300}$ values
of 35.6, 37.1, 37.7, 35.9 and 36.7, respectively. Therefore, we
will again choose KNN/Carhart with $k=8$ as the overall best
model. The total number of model fits is
$10(19+15+6+5\times3)=550$,
with no model requiring more than five random splits of the data.

%
%
\begin{table}
\caption{Simultaneously tuning NN/Burden and KNN/Carhart models for the
PubChem \textit{AID362} assay data. The size and decay values for the NN/Burden
models are defined in Table \protect\ref{tab:nn:burden}; KNN/Carhart
model $k$
has $k$-nearest neighbors. The models surviving after each split are
denoted by a check mark; KNN/Carhart models 2--5 and 6--10 survive the
same number of splits, respectively}
\label{tab:one:search}
\begin{tabular*}{\tablewidth}{@{\extracolsep{4in minus 4in}}lcccccccccccc@{}}
\hline
\multirow{2}{37pt}[-7pt]{\textbf{Number of splits}}
&
\multicolumn{9}{c}{\textbf{NN/Burden model}} &\multicolumn
{3}{c@{}}{\hspace*{-3pt}\textbf{KNN/Carhart model}} \\
[-4pt]
& \multicolumn{9}{c}{\hrulefill} & \multicolumn{3}{c@{}}{\hspace*{-3pt}\hrulefill}\\
& \textbf{1} & \textbf{2} & \textbf{3} & \textbf{4} & \textbf{5}
& \textbf{6} & \textbf{7} & \textbf{8} & \textbf{9} &
\multicolumn{1}{c}{\textbf{1}}
& \multicolumn{1}{c}{\textbf{2--5}} & \multicolumn{1}{c@{}}{\textbf{6--10}} \\
\hline
0 & \checkmark& \checkmark& \checkmark& \checkmark& \checkmark
& \checkmark& \checkmark& \checkmark& \checkmark
&\checkmark& \checkmark& \checkmark\\
1 & & \checkmark& \checkmark& & \checkmark
& \checkmark& & \checkmark& \checkmark
& & \checkmark& \checkmark\\
2 & & \checkmark& & &
& & & &
& & & \checkmark\\
3 & & & & &
& & & &
& & & \checkmark\\
4 & & & & &
& & & &
& & & \checkmark\\
5 & & & & &
& & & &
& & & \checkmark\\
\hline
\end{tabular*}
\end{table}
The second approach, simultaneously tuning and comparing models,
requires less computer time here because the best KNN/Carhart
models outperform all the NN/Burden models, and the latter can be
quickly eliminated. In contrast, the tune-then-compare
implementation spends much computational effort in tuning the
inferior modeling strategy, NN/Burden. It does, however, lead to
an accurate, quantitative assessment of the difference in
predictive performance between NN/Nurden and KNN/Carhart.

\section{Conclusions and discussion}\label{sect:conc}

Throughout we used 10-fold CV, even for $k$-nearest neighbors
where it is computationally straightforward to use $n$-fold
(leave-one-out) CV. We used 10-fold CV for consistency across modeling
methods: $n$-fold CV would be computationally infeasible for the
method of neural networks also considered here and for many other
methods. In addition, $n$-fold CV has well-known limitations.
Theoretically, \citet{Sha1993} showed its inconsistency in model
selection. For applications like the molecular databases in
PubChem, it is also well known that $n$-fold CV can give
over-optimistic estimates of predictive performance if the data
have sets of similar compounds (``twins''). It is easy to predict
one such compound's assay value from its near-analogs.

We illustrated that tuning a model may have a large effect on
predictive performance. We also showed that the variation in CV
performance estimates from one data split to another may
necessitate multiple data splits for reliable comparison of
different sets of tuning parameter values or of different tuned
statistical modeling methods/explanatory variable sets. The
data-adaptive algorithms developed in Sections \ref{sect:alg}
and \ref{sect:compare:methods} attempt to make reliable
comparisons based on enough data splits, but sequentially focus
the computational effort on models with better predictive
performance.

The basic sequential algorithm in Section \ref{sect:alg:split}
uses data splits as a blocking factor, and hence requires at least
two data splits for each candidate model. The variation in
Section \ref{sect:alg:active} uses individual observations as the
blocking factor, and can sometimes eliminate very inferior models
after just one data split and CV analysis. To use observations as
blocks, the performance measure must be an average over
observations. The specialized $h_{300}$ measure appropriate for
the PubChem data set used throughout is of this type, as are more
traditional metrics such as mean squared prediction error in
regression problems or misclassification rate for classification
problems.

The same approach can be applied to tuning a modeling strategy
with respect to user-specified parameters and to comparing tuned
modeling strategies. Simultaneously tuning and comparing will
be computationally efficient relative to nonsequential strategies
if there are many poor modeling strategies that are dominated by other methods.

Parallelization of the algorithms is straightforward, as
regular 10-fold CV is always used for a
specific model and data split.
Thus, with 10 processors, say,
each processor simply performs one of the 10 fits of a single CV analysis.
With more than 10 processors, the 10 fits for each of two or more models
on the same split could be sent to the processors.
Reconciling the results from the parallel computations is fairly trivial;
it is model fitting that dominates computational complexity here.

The proposed data-adaptive CV algorithm is sequential. At each
iteration, a~multiplicity-adjusted statistical test is developed
to eliminate all inferior modeling strategies. An issue not
addressed in this article is how to take account of the multiple
testing across iterations. This is the topic of future study.

We gave an example where two different sets of explanatory
variables were compared. In practice, some statistical models also
have to be ``tuned'' with respect to selection of variables
\textit{within} a given set. This could also be done via our sequential
CV algorithms, at least for a small number of candidate subsets of
variables. Much adaptation would be necessary if there is a
combinatorial explosion of possible subsets, and again this is
future work.\looseness=-1

In practice, some tuning parameters are usually treated as continuous factors,
for example, decay for a neural network. Future study will also include
sequential CV algorithms for continuous factors.


%

%
\printaddresses

\end{document}